\documentclass[conference]{IEEEtran}
\IEEEoverridecommandlockouts

\usepackage{xspace}
\usepackage{color}
\usepackage{xcolor}
\usepackage{amsmath}
\usepackage{url}
\usepackage{booktabs}
\usepackage{multirow}
\usepackage{graphicx}
\usepackage{caption}

\usepackage{pifont}

\newcommand{\cmark}{\ding{51}}
\newcommand{\xmark}{\ding{55}}

\newcommand{\csq}[1]{{\color{blue}#1}}
\usepackage{enumitem}
\newcounter{RZNumberOfComments}
\stepcounter{RZNumberOfComments}
\newcommand{\rz}[1]{\textcolor{teal}{\small \bf [RZ\#\arabic{RZNumberOfComments}\stepcounter{RZNumberOfComments}: #1]}}

\newcommand{\name}{$\sf\small{MetaGuard}$\xspace}

\newcommand{\etc}{\emph{etc.}\xspace}
\newcommand{\ie}{\emph{i.e.,}\xspace}
\newcommand{\eg}{\emph{e.g.,}\xspace}
\newcommand{\etal}{\emph{et al.}\xspace}

\newcommand{\mysubsection}[1]{\vspace{-.07in}\subsection{#1}\vspace{-.02in}}

\newcommand{\revise}[1]{\textcolor{purple}{#1}}

\newcommand{\accessdate}{14-February-2022}

\title{
\huge Towards Zero-trust Security for the Metaverse
}

\author{\IEEEauthorblockN{Ruizhi Cheng, Songqing Chen, and Bo Han}
\IEEEauthorblockA{
George Mason University\\
Email: \{rcheng4,sqchen,bohan\}@gmu.edu}
}

\IEEEoverridecommandlockouts
\IEEEpubid{\makebox[\columnwidth]{\copyright2023 IEEE \hfill}
\hspace{\columnsep}\makebox[\columnwidth]{ }}

\begin{document}
\maketitle
%\fancyhead{}

%\thispagestyle{fancy}
%\pagenumbering{gobble}
%\fancyhead[location]{text}
% Leave Left and Right Header empty.
%\rhead{\thepage}

%\pagestyle{myheadings}
%\markboth{Sundar Iyer}{Sundar Iyer}

% This kind of makes 10pt to 9 pt.
%\begin{small}

%\vspace*{0.3cm}

% \author{\large Ruizhi Cheng, Songqing Chen, and Bo Han}
% \affiliation{%
%   \institution{%Deartpment of Computer Science \\ 
%   George Mason University}
%   %\country{USA}
%   }
%\email{{nwu5,rcheng4,sqchen,bohan}@gmu.edu}

%%%%%%%%%%%%%  ABSTRACT GOES HERE %%%%%%%%%%%%%%
\begin{abstract}
By focusing on immersive interaction among users, the burgeoning Metaverse can be viewed as a natural extension of existing social media.
Similar to traditional online social networks, there are numerous security and privacy issues in the Metaverse (\eg attacks on user authentication and impersonation).
In this paper, we develop a holistic research agenda
for zero-trust user authentication in social virtual reality (VR), an early prototype of the Metaverse. 
Our proposed research includes four concrete steps: investigating biometrics-based authentication that is suitable for continuously authenticating VR users, leveraging federated learning (FL) for protecting user privacy in biometric data, improving the accuracy of continuous VR authentication with multimodal data, and boosting the usability of zero-trust security with adaptive VR authentication.
Our preliminary study demonstrates that conventional FL algorithms are not well suited for biometrics-based authentication of VR users, leading to an accuracy of less than 10\%.
We discuss the root cause of this problem, the associated open challenges, and several future directions for realizing our research vision.

\end{abstract}

%\clearpage
%\section*{Abstract}
%\label{sec:abstract}
%\input{00.abstract}

\section{Introduction}
\label{sec:introduction}
%\bo{
%Discussion of the technical background, previous findings, scientific, and commercial justification for the R\&D. This should include project objectives, outcomes, and intellectual and commercial merit of the proposed work.
%}

Metaverse, with the combination of the prefix ``meta'' (meaning transcending) and the word ``universe'',  has been deemed as a hypothetical next-generation Internet~\cite{cheng2022metaverse}. %\bo{change it to IEEE Network article.}
While there is no consensus on the definition, a narrow depiction of the Metaverse is a universal virtual world for social interaction, by connecting multiple 3D virtual environments via the Internet. % (\ie a network of virtual worlds).
In 2021, Facebook -- now Meta -- planned to invest at least \$10 billion in its extended reality (XR) and Metaverse business for creating novel XR hardware, software, and content. %\footnote{\url{https://bit.ly/3SWdagN}}.

% ~\cite{facebook}.
% \bo{change it to a footnote.}

%Similar to online social networks, 
Undoubtedly, there will be numerous security and privacy issues in the Metaverse, for example,  known attacks on user authentication and impersonation~\cite{pietro2021metaverse}. 
Also, there would be new types of challenges, for instance, defending against immersive attacks~\cite{casey2019immersive} that adjust the location and orientation of users, which may cause collisions with real objects and motion sickness. 
Hence, we should take security and privacy into consideration from day one when designing the Metaverse instead of treating them as an afterthought.

\iffalse
Our long-term research goal is to secure the Metaverse %with a zero-trust privacy-preserving approach, 
by tackling open challenges related to  authentication, %(to identify)
authorization, %(to give permission)
and accounting (3A), management of personal assets such as user-generated content (UGC), protection of interactions between virtual and real-world objects (\eg digital twins~\cite{saddik2018digital}), \etc
%
\fi

In this paper, we focus on user authentication in the Metaverse and argue that it should satisfy the following four key design principles: %principal design objectives:
{\em zero trust, non-intrusive interaction, high reliability, and privacy preservation}. 
Zero trust~\cite{rose2020zero} assumes that no entities (\eg users, transactions, or network traffic) should be always trusted unless verified, as trust itself is a vulnerability. %~\cite{campbell2020beyond}.
Thus, it requires pervasive and continuous deployment of security mechanisms such as authentication and access control~\cite{feng2017continuous}.

\begin{figure}[t]
    \centering
    %\vspace{-.1in}
    \includegraphics[width=0.99\linewidth]{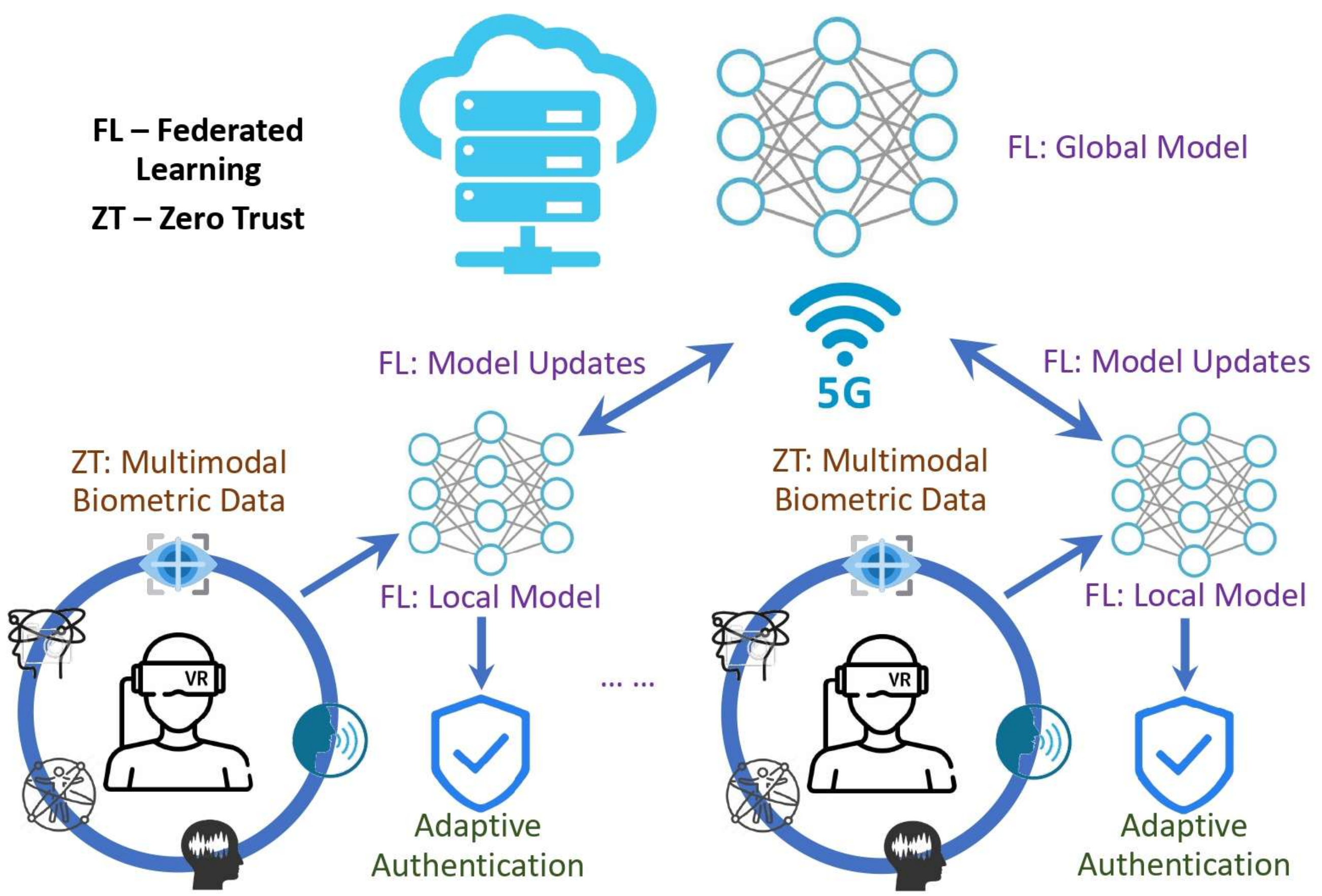}
    %\vspace{-.1in}
    \caption{System architecture of \name.}
    \vspace{-.1in}
    \label{fig:architecture}
\end{figure}

While one-shot authentication on head-mounted displays (HMDs)~\cite{miller2021using} and continuous authentication for mobile devices ~\cite{feng2017continuous} have been extensively investigated, it is still unclear how to {systematically leverage continuous authentication in an immersive environment for effectively supporting zero-trust security}.
Existing studies bear the following limitations.
First, they often utilize a single data source (\eg head movement~\cite{bhalla2021movear}), which may not always be available for continuous authentication and usually fails to guarantee a high accuracy (\ie only around 90\%~\cite{bhalla2021movear}). 
Second, they all benefit from collecting biometric information from users, which is highly sensitive and may potentially lead to severe security and privacy issues.
%~\cite{jain2012biometric}.
%
Third, they rely on intrusive methods (\eg throwing a ball~\cite{miller2021using}) to authenticate users, making them not suitable for continuous authentication.
Lastly, they do not consider the practical challenges of continuous authentication on HMDs (\eg usability and resource utilization). % and energy consumption).
%

%\rz{Swap the order of this item with the previous one?}
%\bo{add a sentence about non-intrusive interaction?}

To address these issues, we explore the emerging security threats and privacy concerns in social virtual reality (VR), an early prototype of the Metaverse~\cite{cheng2022are}, and develop a holistic research agenda to protect it with zero-trust authentication mechanisms.
% \bo{cite our IMC paper here.} and develop a holistic research agenda to protect it with zero-trust authentication mechanisms.
%
In particular, we propose %to build 
a {first-of-its-kind, privacy-preserving continuous and adaptive} authentication framework, dubbed \name (Figure~\ref{fig:architecture}), by employing a {federated-learning (FL) based scheme} that authenticates users with {multimodal biometric data} (\S\ref{sec:research}).
We build a proof-of-concept of \name, focusing on preserving user privacy with FL.
%
%By adopting the fully convolutional networks (FCN)~\cite{long2015fully} as the authentication model and the widely used FedAvg~\cite{mcmahan2017communication} to set up the FL environment, 
Our preliminary study reveals that existing FL models such as FedAvg~\cite{mcmahan2017communication} lead to extremely poor performance since each client holds only positive-label data (\S\ref{sec:preliminary}).
%
%The authentication accuracy is only less than 10\% for a publicly available dataset~\cite{miller2021using} (\S\ref{sec:preliminary}).
%To investigate the challenges of building \name, 
%For this purpose, 
%we first adopt the fully convolutional networks (FCN) ~\cite{long2015fully} as the authentication model, and implement a widely-used FL algorithm, FedAvg~\cite{mcmahan2017communication} to set up the FL environment as an early prototype of \name.
%
%Through experiments, we find that directly integrating FCN  with FedAvg leads to very poor performance.
%for \name.
%
We discuss the root cause of the issues related to this setup and suggest several potential directions for future improvements (\S\ref{sec:discussion}).
% use a widely used FL algorithm -- FedAvg~\cite{mcmahan2017communication}

\section{Background}
\label{sec:background}
\noindent {\bf Zero-trust Security.}
%Zero-trust refers to the assumption by the server that all entities are untrustworthy~\cite{buck2021never}. 
%
Zero trust~\cite{rose2020zero} is a fine-grained security approach that shifts defenses from static perimeters of a protected system to its users, resources, and assets.
It is rooted in the principle of ``never trust, always verify''~\cite{rose2020zero} and eliminates vulnerable permissions and unnecessary access to help service providers better manage and protect identities, applications, and machines across their networks.
%
%Zero-trust is targeted at applications and users, not the network infrastructure~\cite{rose2020zero}.
%
Zero trust requires the system to ensure that all requests are continuously verified before accessing any system asset (\eg continuous user authentication~\cite{feng2017continuous}).
%Thus it requires continuous implementation of security mechanisms for users, such as authentication~\cite{bertino2021zero}.
%
%In this paper, we focus on continuous authentication approach for users in social VR platforms.
%
%An important design goal of the continuous authentication is that it should be non-intrusive.
%
%Otherwise, it will severely impair the user experience~\cite{frank2012touchalytics}.
%
%In this paper, we aim to analyze and solve the problem of continuous authentication for users in social VR to provide a reliable solution to the zero-trust Metaverse.

\vspace{0.05in}
\noindent {\bf Metaverse.} Metaverse could be considered as a large-scale virtual world connected via the Internet.
In recent years, with the rapid development of mobile immersive computing~\cite{han2019mobile}, providing immersive experiences for users via VR headsets has become one of the most crucial goals of the Metaverse~\cite{cheng2022metaverse}.
Therefore, emerging social VR is regarded as a key component of the Metaverse~\cite{cheng2022are}.
In social VR, users are free to explore the virtual scene and socialize with others, such as trading virtual content via the non-fungible token (NFT)~\cite{cheng2022are}.
Thus, a continuous authentication method is required to protect the account and asset security of social VR users.
%\bo{change it to the IMC paper.}
%
%In social VR, since users wear HMDs, traditional user authentication methods, such as PINs, become inconvenient and vulnerable to attacks~\cite{miller2021using}.
%
%This motivates us to utilize other information, such as users' biometric data, for authentication.
%
%In this paper, we propose a privacy-preserving continuous authentication mechanism in social VR to provide a reliable solution to the zero-trust Metaverse.

%\bo{replacing the following with VR authentication? }

\begin{figure}[t]
    \centering
    %\vspace{-.1in}
    \includegraphics[width=0.99\linewidth]{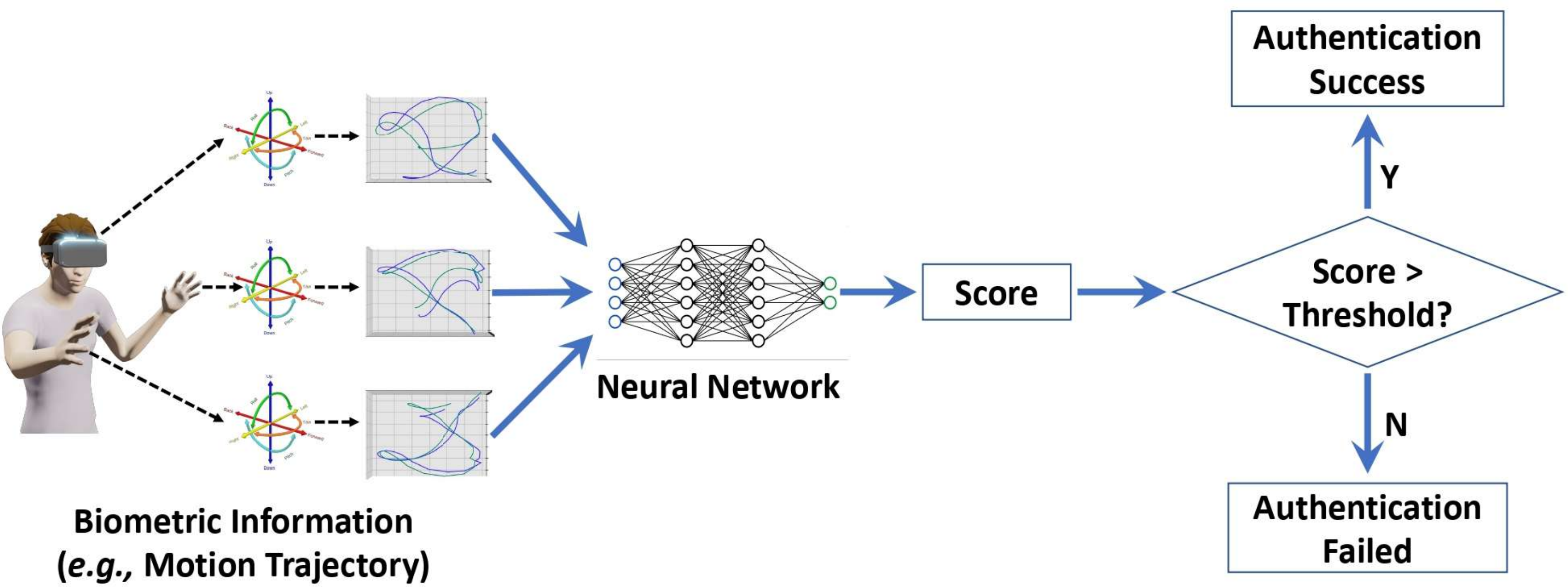}
    %\vspace{-.1in}
    \caption{Biometric-based authentication process in VR.}
    \vspace{-.1in}
    \label{fig:ml_auth}
\end{figure}

\noindent{\bf VR Authentication. }
There are two main approaches for user authentication in VR: knowledge-based and biometric-based.
Since users wear HMDs in VR, knowledge-based methods, such as PINs, become inconvenient and vulnerable to attacks~\cite{miller2021using}.
Therefore, biometric-based authentication has become a more practical option for authenticating VR users~\cite{stephenson2022sok}.
%
%\rz{
Biometric-based authentication in VR often involves training a classification model using machine learning (ML) or deep learning (DL), as shown in Figure~\ref{fig:ml_auth}. %} % to perform the authentication process. 
However, traditional ML/DL models require clients to upload raw data to a server, 
which raises privacy concerns. 
To address this issue, utilizing FL~\cite{yang2019federated} %, a privacy-preserving distributed ML paradigm, 
for VR user authentication is a promising solution.

\section{Threat Models}
\label{sec:threat}
%\review{This section is for Comments 1 and 2. The flow of this section is: First introduces the shoulder surfing attack to show the need for biometric-based authentication. Then introduces the insider attack and human mimicry attack to demonstrate the necessity of continuous authentication. Finally introduces the man-in-the-middle and data poisoning attacks and mentions our proposed personalized multimodal training model can defend this attack.}

In this section, we describe the threats in VR authentication, demonstrating the necessity of continuous biometric-based authentication, which also serve as threat models for  \name.
%\csq{rephrased a bit. threat model is specific to a scheme}

{\bf Shoulder Surfing Attack} refers to a scenario in which an attacker can visually observe a user's interactions with a device during authentication.
This attack is particularly concerning in VR authentication, as users are fully immersed in the virtual environment and may be unaware of their surroundings, making them vulnerable to shoulder-surfing attackers~\cite{stephenson2022sok}.
Unlike knowledge-based authentication methods such as PINs, biometric-based authentication methods could be less susceptible to shoulder-surfing attacks, as they do not lead to direct exposure of the authentication information.
%not lead to the \bo{explicit display of sensitive information}.
%\rz{as they do not require users to input authentication information in the display.}
%\revise{Non-intrusive continuous authentication is promising to resist this type of attack since the authentication process is performed in the background.}
%\csq{not necessary. If users perform certain actions, attacker can also observe and learn. you may tone down a bit like: could be less susceptible... they usually do not lead to direct exposure of the authentication information.}\rz{Is the ``observe and learn'' attack the human mimicry attack?}

{\bf Insider Attack} refers to an individual who has been granted trust, such as a friend or family member, exploits the absence of a device owner to gain unauthorized access to the device or account.
Such attacks pose a significant threat to the security of VR authentication systems, as the insider has already been granted a certain level of access, which they can use to commit account takeover or make unauthorized transactions. 
%
%This type of attack is known as an account takeover attack, where an attacker can commit unauthorized transactions or make in-app purchases. 
%
To defend against this type of attack, a continuous authentication mechanism that can detect and lock the system as soon as the legitimate user is not wearing the device is required~\cite{zhang2017continuous}.
%\rz{Actually I think a better way to defend this attack is to detect if the headset is took off...}

{\bf Human Mimicry Attack} is a form of security threat in which an attacker records a legitimate authentication and subsequently replays it to gain unauthorized access to a device or account. 
In the context of VR authentication, %such attacks could involve 
an attacker could record a user's interactions with the device, such as body movements, and then mimic the recorded movements to impersonate the user and gain access to the device or account.
This type of attack is difficult to detect and prevent, especially in the one-shot authentication~\cite{chen2021user}, as the authentication information used in the replay may %be legitimate and 
appear to be coming from the actual user.
Continuous authentication, with its capability to continuously monitor the user's behavior and detect anomalies, can mitigate such attacks by providing an additional layer of security to authentication.

{\bf Man-in-the-Middle (MitM) Attack} occurs when an attacker intercepts and manipulates the communication between the user and the authentication system.
%
%In VR authentication, the MitM attack can occur during the training phase of a biometric-based (continuous) authentication model, where the attacker alters the biometric data used to train the model. 
%
When training the FL-based authentication model, the attackers can intercept the model sent by the server, train the model using their own biometric data, and then upload the updated model to the server.
This tampering can result in a model that fails to accurately recognize the legitimate user, which enables the attacker to impersonate the user and gain unauthorized access to the device or account.
Thus, one-shot authentication faces a huge risk under this attack.
Even continuous authentication alone does not guarantee security since the attackers can continuously use their data to train the model.
%
%\revise{
Hence, it is imperative to design an additional layer of protection in the continuous authentication scheme to defend against this attack.
{\bf Data Poisoning Attack} is a malicious tactic used by attackers to upload counterfeit data to a ML model with the aim of compromising its performance.
In ML-based classification tasks, such as user authentication, this attack involves the injection of incorrect data for certain labels (users), making them unable to be successfully recognized and classified.
In the context of FL-based authentication systems, if the counterfeit data is irrelevant to other labels, this attack has a limited impact on the authentication performance of others. %since each user only holds data corresponding to their own label.
However, if the counterfeit data resembles that of other users, such an attack can be considered a variant of the human mimicry attack.
\section{Research Agenda}
\label{sec:research}

%\bo{
%What exactly will be done? How will the objectives be met? What are the processes, strategies, methods, studies, testing, trials, or development activities required, and milestones and deliverables for the project? IP and other trade secrets may be protected in this narrative, but sufficient information must be provided for reviewers to assess the feasibility and merit of the scope of the project with the time and resources requested
%}

%The Metaverse mandates an effective approach for protecting both user security and privacy. 
%
%While the Metaverse would be mainly a virtual world, it is closely tied with the physical world. 
%
%For example, transactions conducted in the Metaverse will ultimately be associated with real-world bank accounts.
%
%Thus, authenticating and protecting users' activities in the Metaverse is of paramount importance~\cite{wang2022survey}, which is the main focus of this proposal. 
%
%As shown in the system architecture of \name in Figure~\ref{fig:architecture}, we organize the research agenda into four steps as detailed following.
%building \name into three steps as detailed next.
In this section, we propose a holistic research agenda to empower zero-trust authentication for the Metaverse with four concrete steps by presenting the design of \name.

% propose three tasks in this step.
% We will organize this project of securing the Metaverse into four steps as detailed next.

\vspace{0.05in}
\mysubsection{Benchmarking Biometrics-based Authentication for Metaverse}
\label{sec:investigating}
Early research on authenticating mobile devices~\cite{feng2017continuous} and HMDs~\cite{bhalla2021movear} suggested that collecting physiological (\eg iris, face, fingerprint, voice, and brain wave) or behavioral (\eg typing, touch, hand gesture, and gait) biometrics from users to build the ML model would be an effective way to authenticate them. 
%
%would be an effective way to authenticate them. 
%
%With the increasing complexity of leveraging biometrics for authentication (\eg the shift from physiological to behavioral biometrics), the underlying method has evolved from template-based to machine-learning-assisted authentication.
%
While biometrics-based authentication has been extensively explored, %~\cite{dahia2020continuous,zhang2018survey,ryu2021continuous}, 
it is still unclear how to effectively adapt it for \name to offer privacy-preserving zero-trust security mechanisms. 

To bridge this gap, we plan to systematically investigate different biometric modalities in terms of their {\em usability, %user-side easy to use,
reliability,  %accuracy, 
vulnerability, %spoofing attacks,
collectability, %device-side ease to collect,
sensitivity, and adaptability}. 
For example, continuous authentication in \name should be non-intrusive, which excludes fingerprint as a candidate, although it is more reliable and accurate than behavioral biometrics such as gait and hand gestures.
Also, behavioral biometrics such as head and body motion may be more vulnerable to human mimicry attack than others~\cite{miller2021using}, 
%\bo{cite a paper that is already in the references.} %~\cite{george2019investigating},
whereas physiological biometrics such as face and voice are more sensitive to the environment~\cite{feng2017continuous} and the collection of brain waves is still a challenging task. %~\cite{elor2020shooting}.
Moreover, continuously collecting and processing biometrics, such as gaze movement, may consume more system resources than others.
To this end, we plan to conduct a large-scale user study by designing a VR environment, where participants will engage in various common activities such as walking, reading, and playing games.
While these activities are taking place, we will collect various biometric data, such as body/gaze movements and pupil size, which have been demonstrated to be appropriate for user authentication in VR~\cite{stephenson2022sok}.
By conducting this study, we aim to gain a deep understanding of the efficacy of different biometric modalities for authentication, and to create an open, diverse, and comprehensive dataset for the design and evaluation of \name.

\begin{figure}[t]
 
    \centering
    \includegraphics[width=0.77\columnwidth]{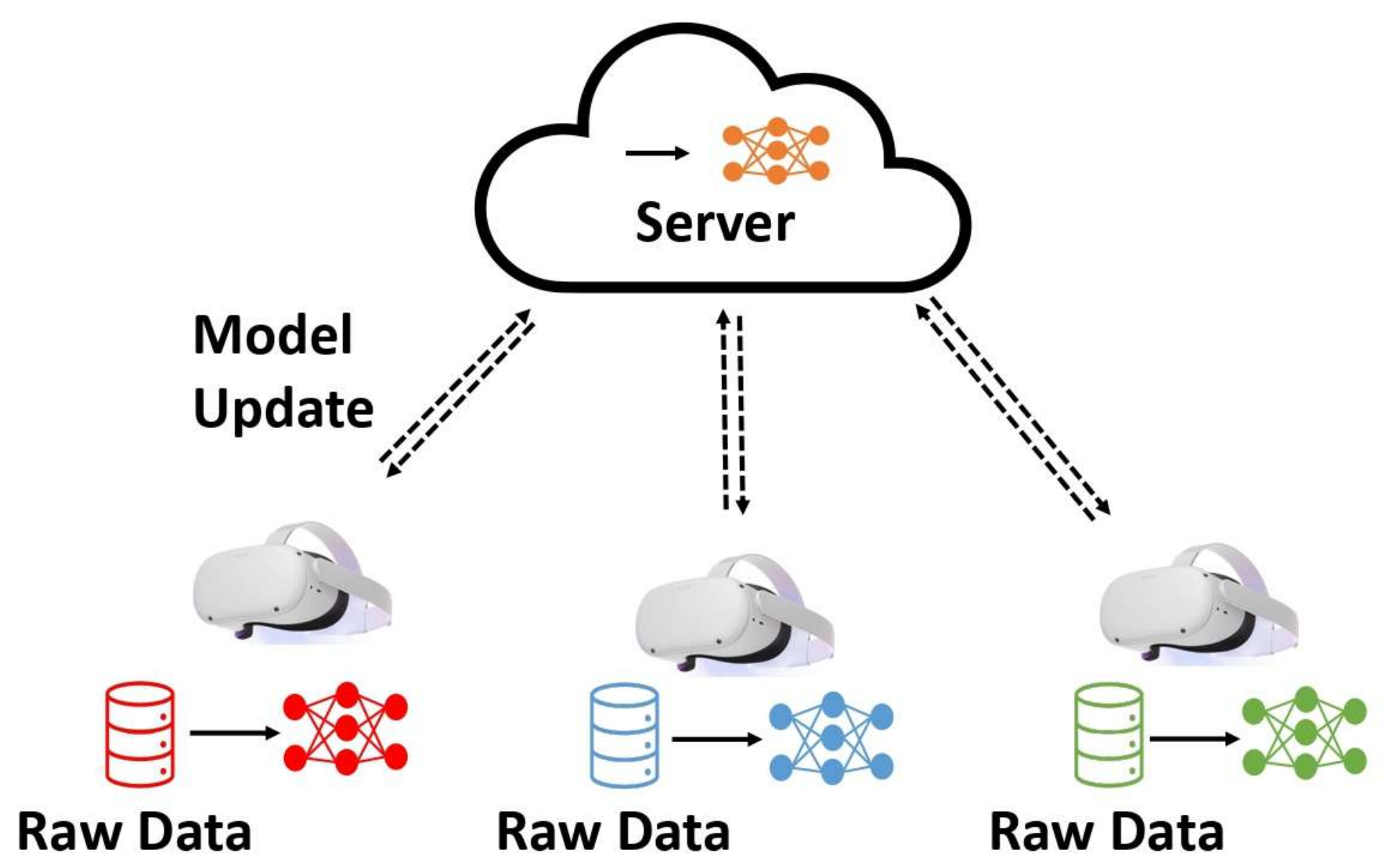}    
    %\vspace{-0.1in}
    \caption{A schematic diagram of training FL-based user authentication model in VR.} 
    \label{fig:FL}
    \vspace{-0.1in}
\end{figure}

\vspace{0.05in}
\mysubsection{Leveraging Federated Learning to Protect User Privacy}
\label{sec:leveraging}
% \noindent {\bf Task 1: Protecting User Privacy with Federated Learning.}
%
Biometrics-based authentication poses severe privacy concerns to users if such a scheme is blindly employed in the Metaverse, as the collected data often carries sensitive and private information. 
If biometric data is misused, leaked, or stolen, users could be endangered. 
To address this concern, we will resort to the recent advances of FL~\cite{yang2019federated}. 
As shown in Figure~\ref{fig:FL}, when training the FL model, the sensitive user data will not be uploaded to a central place (\eg a cloud or edge server).
%\bo{As introduced in \S\ref{sec:background},} the sensitive user data will not be uploaded to a central place (\eg a cloud or edge server) for training the FL model. 
%
%Instead, each user device, such as a smartphone or HMD, trains a model with its raw data locally.
%
%training of a local model with the raw data is conducted on each user device, being a smartphone or HMD.
%
Only the updated gradient of local models will be uploaded to the server for updating the global model, which will, in turn, be used to improve local models. % (Figure~\ref{fig:architecture}). 
This would effectively mitigate the potential risks and protect user privacy. 
%To the best of our knowledge, we are the first to leverage FL for preserving user privacy in biometrics-based VR authentication.

To train a user authentication model with FL, an open challenge is that each client holds only positive-label data, which may lead to scalability issues of the model.
ML-based user authentication is essentially a binary decision problem.
During the training process, the loss function tends to increase the similarity between class embeddings with the same label ($L_{pos}$) and minimize the similarity between those %class embeddings 
with different labels ($L_{neg}$).
%
%Hence, the model can pull apart the class embeddings with different labels in the feature space and obtain the ability to classify and recognize them.
%
While in the FL-based authentication scenario, since each client has only its own data (\ie positive label), the local model cannot optimize $L_{neg}$.
Optimizing only $L_{pos}$ will lead to a trivial solution where all class embeddings collapse into a single point in the feature space~\cite{hosseini2021federated}.
Hence, the model cannot separate feature embeddings among different users, leading the authentication accuracy to decrease as the number of users increases (\S\ref{sec:preliminary}).
To address this problem, we propose to utilize the time-series feature of biometric data and variational recurrent neural network (VRNN)~\cite{chung2015recurrent}, as well as contextual information to design the FL-based authentication model (\S\ref{sec:solution}).

\vspace{0.05in}
\mysubsection{Improving Zero-trust Authentication Accuracy with Personalized Multimodal Biometric Data}
\label{sec:improving}
% \noindent {\bf Task 2: Improving Accuracy of Zero Trust Authentication with Multimodal Biometric Data.}
%
The Metaverse necessitates a reliable zero-trust security framework with highly accurate authentication schemes. % that is to be researched.
Given the increasing stake in the virtual world, conducting traditional authentication methods (\eg password or two-factor authentication) only once at the beginning of the session is no longer sufficient for the Metaverse. 
%
%\revise{
With more users joining the Metaverse, the uniqueness of the authentication features for each user is reducing, making it crucial to continuously collect and train the authentication model. Moreover, as we discuss in \S\ref{sec:investigating}, some biometric information, such as voice, has a dynamic nature that can change over time. Thus, one-shot authentication may not be robust under different environments. %}
%\bo{do you really want to mention body movement here, which is used in MetaFL? }\rz{have deleted ``body movement''}

%\revise{
We point out that a naive continuous authentication approach, which monitors when the headset is taken off from the user's head and performs authentication each time the headset is put on, is unreliable. 
This is because it requires additional sensors to monitor device usage.
Even though it has been integrated into the commercial VR headset, such as Oculus Quest 2, it can be disabled by software-level attacks\footnote{\url{https://bit.ly/3YUUGjt} [accessed on \accessdate]} or physically covered, for example, using tape\footnote{We have confirmed that using tape to cover the Oculus Quest 2's proximity sensor can make it unable to recognize if we take off the headset.}.
Additionally, conducting authentication only after the device is put on is vulnerable to the MitM attack, as it presents a noticeable pattern for the attacker to recognize and intercept the authentication model.
Thus, we need to design a fine-grained continuous authentication mechanism in the Metaverse.
%} \bo{what do you mean by turn on and off?  turning the device off is different from taking it off. }\rz{Have revised}

While prior research has proposed continuous authentication~\cite{feng2017continuous}, a major caveat of these strategies is that, as we mentioned in \S\ref{sec:introduction}, they often rely on a single trait of users.
%or a simple combination of a few features (\eg %keystroke dynamics and behavioral profiling~\cite{saevanee2015continuous}, 
%behavioral biometrics and voice~\cite{peng2017continuous} or hand motion and hold posture~\cite{zhang2021multimodal}) to recognize and authenticate users.
%
However, depending solely on a single biometric modality for authentication in VR may lead to sub-optimal performance~\cite{miller2021using} and be vulnerable to human mimicry attacks~\cite{stephenson2022sok}.
Moreover, the free-interaction nature of the Metaverse presents challenges to the effectiveness of single-modality authentication. 
%
%This, however, is not practical in the Metaverse and may lead to unsatisfactory accuracy. 
%
For example, a user may intermittently walk and sit, leaving a scheme that solely relies on users' gait powerless when they do not walk. 
Similarly, an authentication scheme monitoring users' typing would leave a loophole when they do not behave as expected. 

To this end, we will design a multimodal FL model where multiple modalities of biometric data (\eg pupil size, head motion, body motion, voice, \etc) are collected and fused together for authentication.
Moreover, we plan to propose a personalized FL model~\cite{li2020federated} for each user with a unique combination of modalities that reflects their unique biometric features.
This approach will not only improve authentication accuracy but also effectively defend against the MitM attack.
% the data poisoning attack~\cite{bagdasaryan2020backdoor}.
% %
% \rz{here}
% In addition, our personalized multimodal FL model has the potential to defend against the MitM attack.
%
%If an attacker intercepts the model sent by the server, trains the model using his/her own biometric data, and then uploads the updated model, the server will be able to detect the tampering since the updated model parameters would be hugely different from those from the legitimate user from previous communication rounds.
%
This is because users' multimodal biometric data will present unique features.
%
%Even if the MitM attacker also performs the human mimicry attack,  our model is still capable of defending against such an attack since it is difficult to accurately mimic various biometric information at the same time.
%
Even if MitM attackers also perform the human mimicry attack, our proposed model is still capable of defending against it %such an attack
since precisely mimicking multiple biometric information is challenging.
Furthermore, some biometric data, such as gaze movement, is hard to be mimicked since the user's eyes are covered by the VR headset.
%}~\review{Comment 1} 
%\csq{maybe tone down a bit? check Nvidia Broadcast}
%combination of modalities for the legitimate user would be altered
%To this end, we will design a multimodal approach where multiple modalities of biometric data (\eg eye gaze, head motion, body motion, brain wave, voice, \etc) are collected and fused together for authentication.
%
%For example, the user's keystroke or touch, hand gesture, eye gaze, speaking, gait, etc. can be collected via the inertial unit and sensors in the headset or smartphones while the user is in the Metaverse. This can enable amplification effect as these modalities can complement each other. 
%
However, integrating multiple biometric modalities, particularly when utilizing body movement data, into authentication systems may lead to noise and a decrease in authentication accuracy %\rz{
(\S\ref{sec:preliminary}).
To address this challenge, we propose a density-based within-client modality selection method that chooses the best modality combination for each client (\S\ref{sec:solution}). %}

\vspace{0.05in}
\mysubsection{Boosting Usability of Zero-trust Security with Adaptive Authentication}
\label{sec:boosting}
% \csq{this part is not discussed/experimented, and may be commented out for now}
%\noindent {\bf Task 3: Boosting Usability of Zero Trust Security with Adaptive Authentication.}
%
The Metaverse requires a non-intrusive and easy-to-use authentication that will not impact users' experience while protecting their privacy.
Our proposed FL-based framework with multimodal biometric data as input is expected to provide strong protection to end-users. 
However, HMDs are not well designed to handle computation-intensive tasks such as FL training and continuous data collection, processing, fusion, and inference, which could inadvertently impact user experience, for example, quickly depleting the battery. 

To this end, we will optimize %our proposed zero-trust framework in 
\name with {adaptive and dynamic authentication}.
%~\cite{arias2020survey}.
%
For example, a coarse-grain data collection and authentication frequency can be used when users' behavior does not change.
Otherwise, intensive authentications will be triggered. 
%
%Besides behavioral changes, 
Also, we plan to investigate contextual information that \name can utilize to effectively reduce authentication costs. 
%
%An open challenge we are facing is context manipulation, which makes \name believe the situation does not require authentication.
%
%Another direction that we plan to explore is to utilize edge computing and opportunistically offload some heavy computation to nearby devices such as edge servers and users' smartphones whenever it is feasible. 
An inherent and practical challenge of adaptive authentication is to balance the trade-off between reliability, usability, and resource consumption.
For example, in traditional FL algorithms such as FedAvg~\cite{mcmahan2017communication}, the selection weights of all clients are the same by default.
%initially. %\bo{by default}.
%
In each communication round, the server randomly selects clients to participate in the training.
However, clients may present heterogeneous resource and data issues, affecting the training efficiency and model performance~\cite{diao2021heterofl}. 
%\bo{cite a paper that is already in the references.} %~\cite{nishio2019client,cho2020client}. % due to the different available resources and data quality.
%
% Second, when training the FL model, the weights of all client nodes are the same by default.
% %
% That means, in each communication round, the server randomly selects some clients or chooses all the clients to participate in the training.
%
%When training the FL model, 
%If the selected clients have poor data quality or insufficient available resources, they may affect the training efficiency and model performance~\cite{diao2021heterofl}.
%
To address this issue, we need to design an adaptive client-selection algorithm that chooses the appropriate clients to jointly train the global model in each %FL communication 
round. %, to solve the heterogeneous resources and data issues.
%of the client.
%
A unique challenge of \name is that we cannot simply drop a client for any reason, as everyone should be continuously authenticated in the Metaverse.

\section{Preliminary Results}
\label{sec:preliminary}
To better understand the technical challenges of \name, we build a proof-of-concept that partially implements our proposed research in \S\ref{sec:leveraging} %\rz{
and \S\ref{sec:improving}.
In this section, we present our preliminary results of FL-based VR authentication that shed light on realizing the fully-fledged \name.

%In this section, w
%We first implement a conventional FL algorithm on a publicly available biometrics-based VR authentication dataset as a proof-of-concept for \name.
%an early prototype of \name.
%
%Although we only partially implement the research plan in \S\ref{sec:leveraging}, it can provide us with some initial understanding of the challenges when building the fully-fledged \name and shed some light on our future work.
%
%As a first step, we implement a widely adopted FL algorithm~\cite{mcmahan2017communication} and compare its performance with three non-privacy-preserving deep learning methods.
%on the biometrics-based VR authentication task. 
%
%We then discuss the reason for the poor performance of this naive setup and propose some potential solutions. %for building the fully-fledged \name.
% Then we discuss the the reason of the bad performance of FedAvg, and discuses some potential solutions.

% \vspace{0.05in}
% \mysubsection{Experimental Settings and Results}
% \label{sec:experimental}

 \vspace{0.05in}
\noindent {\bf Dataset.} We use the dataset released by Miller~\etal~\cite{miller2021using} for our initial experiments.
%
%, which is the largest biometrics-based VR authentication dataset in terms of the number of users.
%
This dataset was collected with an application of throwing a ball in VR.
It contains the trajectories of 41 users throwing the ball using the Oculus Quest VR headset %\rz{
and has 6 modalities, including the position and orientation of the headset and both controllers.
Each user threw the ball 10 times per day for two days.
We use the first-day trajectories for enrollment (training set) and the second-day trajectories for authentication (test set).
%
%Note that requiring ball-throwing for authentication is an intrusive authentication approach that will not work for continuous authentication.
Note that leveraging ball-throwing for authentication is intrusive and may not be suitable for continuous authentication.
However, given that this is the largest publicly available biometrics-based VR authentication dataset, in terms of the number of users, we believe it can help us gain some initial insights into %the challenges of 
designing privacy-preserving continuous authentication for the Metaverse.

\vspace{0.05in}
\noindent {\bf Implementation Details.} %
We implement three well-known deep learning (DL) models that have been demonstrated %by previous work~\cite{miller2021using}
to achieve high accuracy in VR authentication tasks~\cite{miller2021using}: siamese neural network, FCN, and ResNet.
%\revise{
FCN and ResNet are based on a benchmark of DL-based classification~\cite{ismail2019deep}. 
Both models have a similar architecture, consisting of three convolutional layers and ending with a dense layer using softmax as the activation function. 
The output of these two models is a set of $N$ scores, where $N$ is the number of users, representing the probability of each candidate being the authenticated user.
The candidate with the highest score is then identified as the authenticated user.
The siamese network has the same limb as the FCN but does not have a dense layer.
%
%It receives \bo{two input trajectories}
%\rz{
It receives two pieces of data %from either the same user or different users
as input (\eg enrollment and authentication motion trajectories), %},
calculates their Euclidean distance, and outputs the distance.
The user ID of the enrollment trajectory with the closest distance to the authentication trajectory is the identified ID.
%}~\review{Comment 4}
%
%For FCN and ResNet, we implement the network architecture based on a benchmark of DL-based classification~\cite{ismail2019deep}.
%
%\rz{detail the architecture so that we can delete the above reference}
%
%These two models end with a dense layer using softmax as the activation function.
%
%The models' output with the maximum value is the identified user ID.
%
% We consider the first-day trajectories as the enrollment trajectory and the second-day trajectories as the authentication trajectory for these two models.
%
%The siamese network has the same limb as the FCN.
%
%The only difference between them % siamese network and the FCN 
%is that the siamese network receives two input trajectories and outputs their Euclidean distance.
%of these two trajectories.
%
%The user ID of the enrollment trajectory with the closest distance to the authentication trajectory is the identified ID.
%We use the user ID of the enrollment trajectory with the closest distance to the authentication trajectory as the identified user ID.
%
%We split the user to training user and testing user based on the setting of the previous work~\cite{miller2021using}, and we use 5-fold cross validation to evaluate the siamese network model.
%
%Note that %in the above three model setups, 
The server trains the above three models with all users' data (\ie trajectories of headset and controllers), compromising their privacy.

%
%to know the user's registration data and identification data to train the model and perform authentication, so the user's privacy is leaked. 
%
%To protect the user's privacy, we use federal learning to train the authentication model. 
Next, we set up and train the FL model that ensures each client utilizes only its own data and does not share it with others, including the server.
We leverage a widely used FL algorithm -- FedAvg~\cite{mcmahan2017communication} -- to enable the server and clients to jointly train the model. 
The process of training the FL model via the FedAvg algorithm is as follows: %\rz{Figure?}:
\begin{enumerate}[label=(\alph*)]
    \item \label{step1} In the $k^{th}$ communication round, the server sends the class embedding $W_{k}$ and model parameters $\theta_{k}$ of the global model to all clients.
    
    \item \label{step2} The $t^{th}$ client updates ($W_{k}$, $\theta_{k}$) to ($W_{k}^{t}$, $\theta_{k}^{t}$) based on its local data and the loss function.
    
    \item The server receives the updated ($W_{k}^{t}$, $\theta_{k}^{t}$) from all clients and updates the ($W_{k+1}$, $\theta_{k+1}$) of the global model by taking a weighted average of ($W_{k}^{t}$, $\theta_{k}^{t}$).
    
    \item \label{step4} The server transmits the ($W_{k+1}$, $\theta_{t+1}$) to all clients and keeps repeating steps~\ref{step2}--~\ref{step4} until the global model is converged.
\end{enumerate}

Based on the above algorithm, the authentication model would not leak user privacy.
We use the FCN as the network architecture of the global model for FL, namely FedAvg + FCN.
% 
%We do not use the siamese network or ResNet because training siamese network needs to have both positive samples and negetive samples~\cite{koch2015siamese}, while each client holds only positive samples (its own data), and the accruacy of FCN is better than ResNet, as shown in Table~\ref{table:diff_model_acc}.
%
%We use all captured trajectories (position and orientation of the headset and both controllers) to train the model given that it can get the best performance in this dataset~\cite{miller2021using}.
%
%This also demonstrates the importance of using multimodal data to build the authentication model (\S\ref{sec:Designing}).
%
%We perform Z-score normalization~\cite{patro2015normalization} to
%normalize~\csq{how is data normalized?} 
%
%all data before training and 
We use the Adam optimizer with a learning rate of 0.001 for training.
For the three non-privacy-preserving models, each training runs 2,000 epochs.
%we train 2,000 epochs for each training.
%
For the FedAvg + FCN model, we train it with 100 communication rounds and let all clients participate in the training during each round.
For each training, we save the model with the best authentication accuracy in the test set.
%We save the model with the maximum accuracy in the test set for each training.
%
%We make sure that the model converged at each training and save the model with the best performance for evaluation.
%
% for 100 epochs (communication rounds in the FL setting) and save the model with the best performance for evaluation. \csq{why 100 epochs? does not seem to be long. Does 100 get to a stable result?}
%
We train each model five times and report the average accuracy. % as the model performance.
%\csq{why 100 epochs? does not seem to be long. Does 100 get to a stable result?}\rz{has rewrote the training details}

%\csq{train? test?}
% \rz{should be train? The results of each training model may be different, so I train the model five times and take the average of the five results as the performance of this model}
% \csq{ok}

\begin{table}[t]
  
    \centering
    \begin{tabular}{c|c|c}
        Model&Privacy Preservation&Accuracy\\
        \cline{1-3}
        Siamese Network&\xmark&90.2\%\\
        \cline{1-3}
        FCN&\xmark&89.3\%\\
        \cline{1-3}
        ResNet&\xmark&87.2\%\\
        \cline{1-3}
        FedAvg + FCN&\cmark&6.34\%\\
    \end{tabular}
    %\vspace{-0.05in}
    \caption{Accuracy of different authentication models. }
    \label{table:diff_model_acc}
    %\vspace{-0.1in}
\end{table}

\vspace{0.05in}
\noindent {\bf Results.}
% \mysubsection{Results}
% \label{sec:result}
Table~\ref{table:diff_model_acc} shows the accuracy of different models when using all 41 users' trajectories with all 6 modalities, which leads to the following two observations.
First, %we observe that 
on this intrusive authentication dataset, 
the accuracy of the best non-privacy-preserving model, the siamese network, is only $\sim$90\%.
However, as we discussed in \S\ref{sec:investigating}, to perform zero-trust continuous authentication, \name must utilize %be built based on the
non-intrusive approaches, whose accuracy may be even lower than the intrusive solutions.
%\csq{the reason is not intuitive -- add a short sentence to explain why?}\rz{done}
%
For example, Pfeuffer~\etal~\cite{pfeuffer2019behavioural} leveraged users' gait movement to perform authentication, and the accuracy is only $\sim$50\%.
%
% a previous work using users' gait movement during walking to perform authentication had an accuracy of only $\sim$50\%~\cite{pfeuffer2019behavioural}.
%
One possible reason is that it is more difficult to differentiate the features of user data in non-intrusive methods, which introduces %the first 
a key challenge: % of \name: % needed to be addressed: 
%\textit{How to improve the accuracy of non-intrusive authentication to ensure the effectiveness of continuous authentication?}
\textit{How to improve the accuracy of continuous authentication based on non-intrusive methods to ensure the effectiveness of \name?}

\begin{figure}[t]
 
    \centering
    \includegraphics[width=0.77\columnwidth]{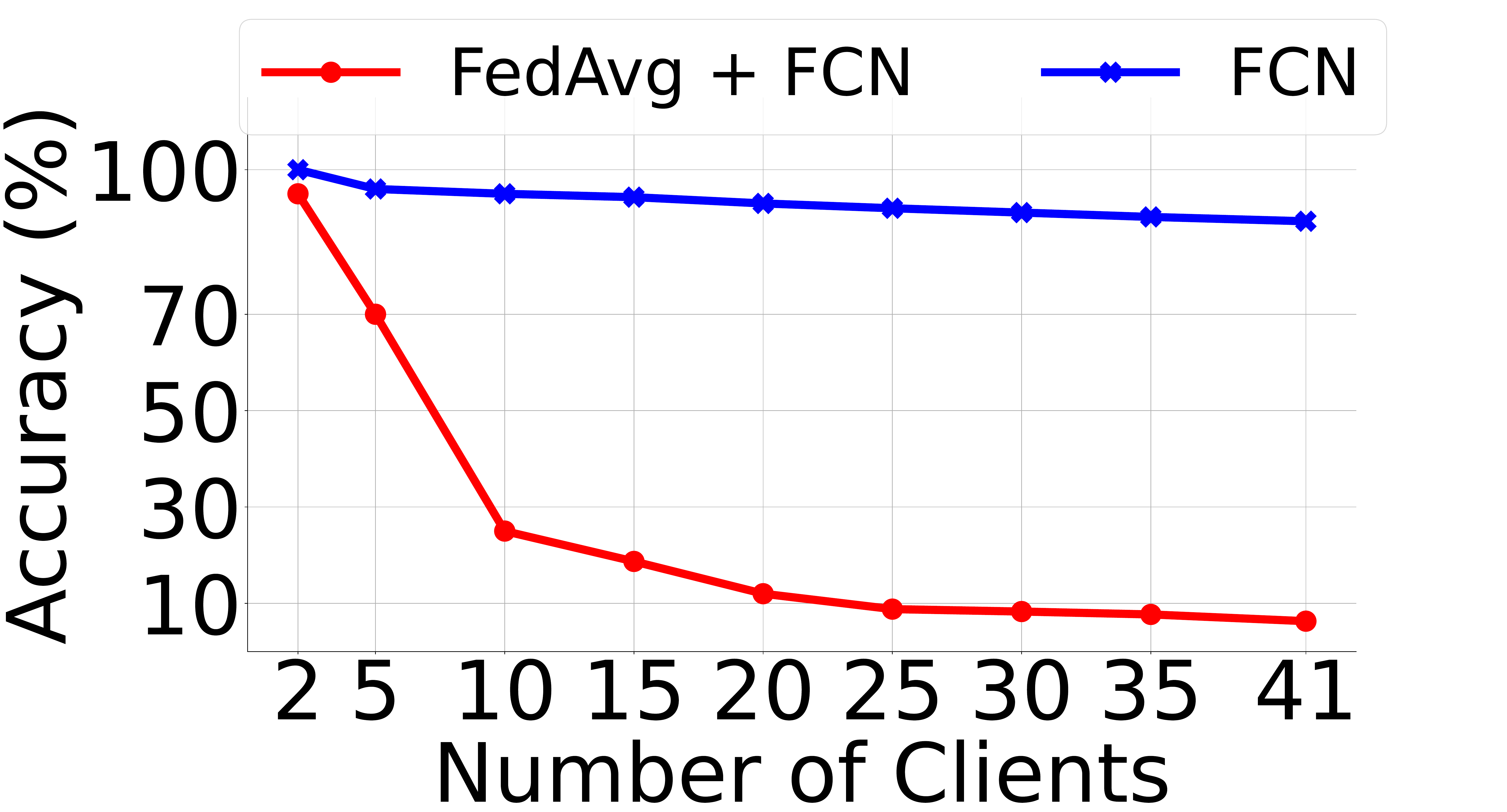}    
    %\vspace{-0.1in}
    \caption{Impact of the number of clients on the FedAvg + FCN and FCN models.} 
    \label{fig:fedavg}
    \vspace{-0.1in}
\end{figure}

Second, %we observe that 
the FedAvg + FCN method can achieve an accuracy of only 6.34\%, while all non-privacy-preserving models have %can lead to 
higher than 87\% accuracy.
The reason may be that %is might be because 
%These results suggest that the 
FedAvg + FCN cannot distinguish the features of most users, resulting in a low authentication accuracy when the number of users is large.
To understand the impact of the number of users on the authentication performance of FedAvg + FCN, we train the FCN and FedAvg + FCN models with different numbers of users using all 6 modalities and show the result in Figure~\ref{fig:fedavg}.
As we can see, the accuracy of the FCN model only slightly drops when the number of users increases.
However, FedAvg + FCN is significantly affected by the number of users.
When there are only two users, it can achieve 95\% accuracy.
However, when the number of users increases to 10, the accuracy reduces to $\textless$30\%; when the number of users further increases to 25, the accuracy is under 10\%.
%
% Whereas when there are 10 clients, the accuracy drops to less than 30\%, and when there are 25 clients, the accuracy is under 10\%.
%
%However, g
Given that the Metaverse will accommodate tens of thousands of users from all over the world~\cite{cheng2022metaverse}, the user authentication model must be scalable and robust.
Thus, it leads to another key challenge for \name: % needed to be addressed:
\textit{How to ensure the accuracy of \name remains stable when the number of users increases?}

\begin{figure}[t]
 
    \centering
    \includegraphics[width=0.77\columnwidth]{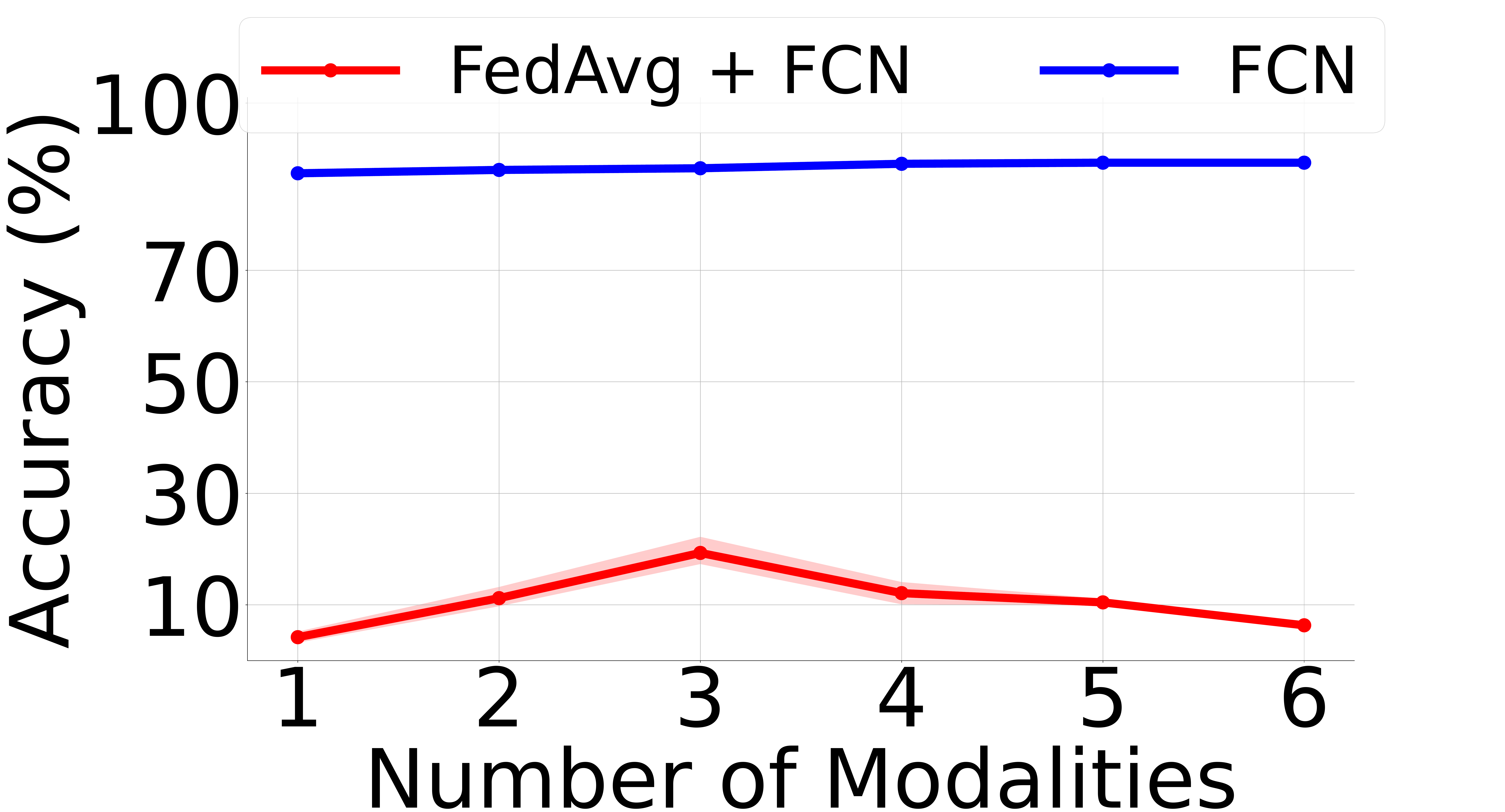}    
    %\vspace{-0.1in}
    \caption{The average accuracy of FedAvg + FCN and FCN models using different numbers of modalities for 41 users. The bands represent 95\% confidence intervals. For each number of modalities $x$, it has $\binom 6x$ modality combinations.} 
    \label{fig:diff_modality}
    \vspace{-0.1in}
\end{figure}

%\rz{
Next, in order to understand the impact of the number of modalities on authentication performance, we train the FCN and FedAvg + FCN models with different modality combinations. 
Among the 6 modalities, there are a total of $\binom 61 + \cdots + \binom 66 = 63$ modality combinations. 
The result is shown in Figure~\ref{fig:diff_modality}. %\bo{for 41 users?}\rz{yes}
We observe that the accuracy of FCN is stable with different numbesr of modalities.
However, using fewer modalities may result in higher accuracy for the FedAvg + FCN model.
For example, when using all 6 modalities, the accuracy is merely 6.34\%.
However, when using only 3 modalities, the average accuracy is 19.32\%.
Moreover, the best modality combination may be different for different users.
During the training phase, we record the modality combinations with the highest local accuracy for each of the 41 users and observe that there are 30 unique modality combinations among the recordings. 
These findings demonstrate the importance of personalized modality selection for improving the performance of FL-based authentication models in VR.
%}

\section{Discussion and Future Directions}
\label{sec:discussion}
\vspace{0.05in}
\mysubsection{Discussion}
%\label{sec:discussion}
% The preliminary results indicate that our naive FL setup is not suitable for the privacy-preserving user authentication system in the Metaverse.
%
%the traditional FL algorithm should not be directly applied to a privacy-preserving user authentication system for the Metaverse. %with many users. 
%
% \rz{our proposed solution focuses on the model, instead of FL algorithm. looks like we say `` the traditional FL algorithm should not be ...'' is not precise}
%
%We point out that the root cause of this phenomenon is 
%the poor performance of the traditional FL algorithm in this scenario is
% This is because each user holds \textit{only} its own data, which poses the following problem.
% %
% During the training process of FedAvg (\S\ref{sec:preliminary}), the loss function used in step~\ref{step2} is supposed to increase the similarity between the instance embedding and the positive class embedding ($L_{pos}$), and minimize the similarity between the instance embedding and the negative class embedding ($L_{neg}$).
% %
% While in the user authentication scenario, the $t^{th}$ client has only its positive class embedding $W_{t}$ and thus cannot compute $L_{neg}$.
% %
% Optimizing only $L_{pos}$ will cause all class embeddings to collapse into a single point in the feature space~\cite{hosseini2021federated}.
% %
% Hence, the model cannot separate the feature embedding among different users, leading to a sharp decrease in authentication accuracy as the number of users increases.

% and the authentication accuracy decreases sharply as the number of users increases.
The preliminary results validate our identified challenges % demonstrated problem 
in \S\ref{sec:leveraging} %, which is the 
and demonstrate that the FL-based authentication model may not scale due to the fact that each user holds only positive label data.
%
% the key challenge for FL-based authentication in VR is to alleviate the problem that the model does not scale due to each client holding only positive label data.
%
Next, we discuss the solutions to this problem proposed in existing works and why they do not apply to VR authentication.
%Although there has been no work studying how to leverage FL to preserve user privacy in VR authentication, there is already some work proposing potential solutions in other research community.
%
For example, %Two 
recent work proposed to address this issue by improving the model update process.
%
%the problem of federated learning with only positive labels~\cite{hosseini2021federated,yu2020federated}.
%
%FedAws~\cite{yu2020federated} is a method where the client still optimizes only $L_{pos}$, and the server computes the \bo{average client-trained model} and updates it with an approximate $L_{neg}$.
%
%\rz{
FedAwS~\cite{yu2020federated} is a method where the server, in addition to averaging the uploaded gradients, performs a geometric regularization to ensure users' embeddings are separated by a pre-defined margin. %}
FedUV~\cite{hosseini2021federated} uses an error-correction code to generate a unique secret vector for each user to update the model. %andperform authentication.
%
%Both of the above approaches still consider the case where the client has only positive labels and do not introduce additional data.
%, thus not causing significant extra computational and communication overhead.
%
% \rz{here}
%However, neither of them is as accurate as the state-of-the-art non-privacy-preserving deep learning models on the large-scale datasets they use.
%
%One of their main limitations is their reliance on pre-defined parameters, such as the margin in FedAws and the error-correction code in FedUV, to model the dissimilarity between clients in the feature space. 
%
However, these models do not consider the quality of local data (\ie each client simply uses all of its local data for training). 
This can be problematic, as low-quality local data, such as noise, can negatively affect the accuracy of the trained model.
%
%
%Thus, to improve the authentication performance in \name, it is crucial to consider the quality of local data in order to extract meaningful features for training.

%
%However, they can only obtain $\sim$90\% accuracy on large-scale image classification datasets, such as CIFAR-10~\cite{krizhevsky2009learning}, for which the state-of-the-art models have achieve >99\% accuracy~\cite{cifar10sota}.

%Two 
Other recent work %from the 
on face recognition %area have 
reported an accuracy comparable to the state-of-the-art non-privacy-preserving deep learning models.
For example, FedFace~\cite{aggarwal2021fedface} proposes to first deploy a feature extractor on the server that has been pre-trained on a large-scale dataset. 
This feature extractor is then used to initiate the class embedding for each client to ensure that it lies inside the generated space of the feature extractor.
FedFR~\cite{liu2022fedfr} is motivated by FedFace and introduces extra data created from the public datasets that have low similarity to the data held by the client for computing its $L_{neg}$.
However, both solutions have the prerequisite of %publicly available
large-scale datasets.
%for depolying a pre-trained model in the server. 
%
This is feasible for %in the 
face recognition %area 
because there are numerous public large-scale datasets available, such as the DigiFace1M\footnote{\url{https://github.com/microsoft/DigiFace1M} [accessed on \accessdate]}, which has 1.2M images obtained from 110K individuals. 
%
%MS-Celeb-1M~\cite{guo2016ms}, which has 10M images obtained from 10K individuals. 
%
On the other hand, biometric datasets, especially for VR headsets, are limited in their scale.
For example, as we introduced before, the dataset released by Miller~\etal~\cite{miller2021using}, which is the largest among the public biometrics-based VR authentication datasets, has only 41 users.

The root cause for this disparity in size between the face recognition and biometrics-based authentication datasets is that face images are relatively easy to collect, requiring only devices with cameras, and many people voluntarily share their photos on the Internet.
Whereas collecting users' biometric data on HMDs requires additional setup for sensors, and users may be reluctant to share biometric data because it could contain sensitive information.
Therefore, it is probably unrealistic to expect the publicly available biometric dataset for VR authentication %on HMDs 
to be as large as the face recognition datasets.
Nevertheless, the size of the dataset is directly related to the accuracy of the trained models.
%
%\rz{
Moreover, as discussed in \ref{sec:improving}, capturing the user's whole face poses challenges in VR since the headset %fully 
encloses half of it.
%}
%
Hence, it may not be feasible to apply the methods for face recognition directly to VR authentication.

\vspace{0.05in}
\mysubsection{Future Directions}
\label{sec:solution}

%\rz{
To improve the performance of the authentication model while preserving user privacy, we first propose a density-based within-client modality selection algorithm.
Our approach is motivated by the observation that the optimal modality combinations may differ among users (\S\ref{sec:preliminary}) and %\bo{that contributing modalities} \rz{
that modalities beneficial for authentication are typically well-clustered~\cite{oza2021federated}.
To select the optimal modality combination for each user, we plan to leverage the density-based clustering algorithm, such as mean shift and OPTICS~\cite{xu2005survey}, for reducing noise and calculating the density for each modality combination.
First, after applying the clustering algorithm on each modality combination, the clustered group with only a single point will be considered as noise and removed.
Second, by calculating the distance between the centroids of each clustered group and the center of all data, we will obtain the density of each modality combination.
In this manner, each client can select the modality combination with the highest density for training the authentication model.
%}

%\rz{
Next, we exploit the unique features of the biometric data collected on VR headsets to improve the scalability of \name. %}
%To improve the accuracy of the authentication model while preserving user privacy, we need to exploit the unique features of the biometric data collected on VR headsets.
%
We notice an important fact %feature of them is 
that since all data are captured from the continuous activities of the user, they may present time-series features, such as lag features and rolling window features~\cite{hamilton2020time}, which are not available in other authentication methods, for example, those that benefit from face recognition.
To leverage the time-series features of biometric data on VR headsets, we can perform time-series analysis (TSA)~\cite{hamilton2020time} to
%, such as using auto regressive (AR) and moving average (MA) models
%to analyze its time-series feature and
extract data with salient features for authentication.
For example, by considering the excellent performance of the recurrent neural network (RNN), such as long short-term memory (LSTM)~\cite{hochreiter1997long}, in time-series classification tasks, we can train the RNN-based model to improve the accuracy of non-intrusive authentication.
%
% each user's data is time-series data. 
% %
% Therefore, it is very likely that these data possess some time-series features, such as that are not present in the face image datasets.
%

\iffalse
Although there is no existing study %no work has focused 
on analyzing these features, previous work has hinted at this possibility.
%
For example, Ajit~\etal~\cite{ajit2019combining} found that in the VR authentication with ball throwing, removing the latter portion of the trajectory led to better performance.
%
It may be because the data in the tail part of the ball throwing is not time-series correlated with the previous trajectory.
%
%Therefore we can use TSA to remove this irrelevant part of data and improve the data quality.
\fi

%Motivated by 

%
% To the best of our knowledge, this is the first research plan to use a combination of TSA and RNN to improve the performance of VR authentication.

%Next, for privacy protection, we plan to design a LSTM-based FL model without the scalability issue for VR authentication.
%deploy our RNN-based authentication model under FL settings.
%under FL settings for authentication while protecting user privacy.
%
In particular, we plan to design an LSTM-based FL model for preserving privacy in VR authentication.
In contrast to FCN, training LSTM on the client is less subjective to the negative impact of having only positive labels, mitigating the scalability issue, as shown in Figure~\ref{fig:fedavg}. %\bo{what is this issue? }
The reason is that when updating the gradient of the loss function, LSTM considers only the current time-series data instead of comparing it with other data. %~\cite{greff2016lstm}.
%
%Nevertheless, 
However, since users still have only their own time-series data, if we use the traditional softmax loss-based training approach, this may lead to the model overconfidence problem~\cite{guo2017calibration}.
%
%Because, 
Assuming that there are $N$ users, in the softmax loss-based training model, the last layer of the model is a dense layer with $N$ units, and the output of the model is the recognition probability of these users.
With this training model, the user with the highest %recognition
probability is considered the recognized user.
However, since each user in our case does not have other users' data, the local model updated by a given user does not have any recognition capability of others, %resulting in 
making the output of the other $N-1$ probabilities %being 
meaningless. 
In this secenairo, model's false positive rate (FPR) will increase as $N$, leading to potential scalability issues~\cite{guo2017calibration}.%\bo{not the real problem?}
%since each client only trains its own time-series data, if we use the traditional softmax loss-based training approach, whose output is the probability over N users (N is the number of total users), the client class embeddings still cannot be separated well in the feature space, which may lead to the model overconfidence problem\cite{guo2017calibration}.

To address this problem, we can resort to the VRNN model~\cite{chung2015recurrent} for privacy-preserving authentication.
% \rz{Figure?}
%
VRNN integrates the variational autoencoder (VAE)~\cite{kingma2019introduction} model into the RNN model.
%
%VAE is an encoder-decoder-based generative model that is widely used in reconstruction-based anomaly detection~\cite{pang2021deep}.
%, which is a kind of encoder-decoder-based generative model.
%
In VRNN, RNN is the encoder, VAE samples the output of RNN to reconstruct the time series similar to the input data, and the loss function is the reconstruction error.
%of the resconteced data and the input data.
%
%It has a generator for generating the similar sample of the input data, and its output is the similarity between these two samples.
%
VAE is based on the assumption that it can generate a sample with high similarity to the input data if its label has been trained before.
Conversely, if the model cannot generate a sufficiently similar sample, this input data is considered anomalous. 
%
%When training the VRNN model, the loss function of the model is the reconstruction error of the time series data.
%
Hence, when training VRNN in the FL setting, each user-trained model focuses on minimizing the error of reconstructing its time series.
%, which enlarges the distance of class embedding among users in the feature space.
%
After training, the converged VRNN model will be able to reconstruct the time series of all users involved in the training.
In the authentication stage, the model will make a judgment based on the reconstruction error.
%
%, when a user tries to conduct authentication, the model will make the judgment based on the reconstruction error.
%similarity of the reconstructed data and the data provided by the user. 
%
It will consider the authentication successful when the reconstruction error is lower than the pre-defined threshold, instead of directly using the aforementioned probability.
As a result, this approach increases the confidence level and stability of the authentication model.

%\revise{
To further improve the scalability of our system, we plan to utilize contextual information in VR to conduct authentication.
For example, there is a large number of user-generated content (UGC) in the Metaverse.
Thus, we can design a fine-grained data processing method, which categorizes time-series data based on recent user interactions with the UGC.
During authentication, the user will be verified using the historical information of their interactions with the UGC.
%
%This approach can enhance the authentication performance by considering only the most relevant historical data.
%
Given that users may exhibit distinct patterns in their interactions with different UGC, this approach can help the model learn more useful representations of the user's behavior by considering mainly the most relevant historical data (\ie the interaction with the UGC), enhancing the authentication performance. %}
% \bo{still not clear}
%\csq{sounds like weight of data. Maybe leave as is since we don't want to say too explicit before our paper is published}

%\revise{
%\rz{
We finally discuss three potential directions for further improving the practicality of \name. %}
First, training the VRNN-based FL model on VR headsets may consume substantial resources~\cite{hard2018federated}.
As we discussed in \S\ref{sec:boosting}, clients with insufficient available resources may affect the convergence speed of the global model, which motivates us to design a client-selection algorithm to accelerate the convergence of the model and reduce the communication and computation overhead, without sacrificing the authentication accuracy of \name.
%
%\rz{
Second, continuous authentication necessitates non-intrusive data collection.
However, given that the data collected in this way may exhibit random patterns, especially for behavioral-based biometrics, training the authentication model using raw data may result in poor authentication performance~\cite{pfeuffer2019behavioural}.
Applying signal processing algorithms, such as fast Fourier transform (FFT), to extract features from the raw data and use them to train the authentication model can potentially improve authentication performance~\cite{patel2016continuous}. %}
%
%Utilizing contextual information is a promising solution for enhancing the performance of continuous authentication \rz{by providing a source of unique and relevant features that can help the authentication model better distinguish between different users.}~\cite{patel2016continuous}. \bo{why and how?} \rz{maybe delete ``that help...'' since this sentence is too long?} \bo{is the second one redundant with what has been said in the previous paragraph? }
%
Third, although FL can prevent the exposure of raw user data during the learning process, the model updates communicated in \name may still reveal some user information~\cite{kairouz2021advances}.
In the future, we will integrate other privacy-preserving techniques, such as homomorphic encryption~\cite{zhang2020batchcrypt}, in \name, and investigate the trade-off between privacy and utility.

\section{Conclusion}
\label{sec:conclusion}
In this position paper, we presented a holistic research agenda for securing the Metaverse through a zero-trust continuous authentication framework.
We first analyzed the challenges of conducting privacy-preserving continuous authentication in the context of social VR (\ie a prototype of the Metaverse).
We then proposed \name, a first-of-its-kind, FL-based adaptive and continuous authentication framework for VR users by leveraging multimodal biometric data.
%
%We propose to leverage FL, a privacy-preserving machine learning approach, for continuous authentication of VR users based on multimodal biometric data.
%
%To understand the challenges of building \name, we have implemented the conventional FL algorithm as our preliminary work and found that it is not suitable for \name.
%on a biometric-based VR authentication dataset.
Our preliminary study through a proof-of-concept of \name revealed that blindly applying FL to VR authentication will lead to an accuracy of lower than 10\%.
%
%We find that the traditional FL algorithm does not work for \name, and
We finally discussed the root cause of these problems, which shed light on future improvements.
% and discuss our preliminay result
% the challenges and potential solutions for leveraging FL on continuous authentication in VR.
We hope our study can inspire more research to realize the grand vision of the Metaverse by tackling its security and privacy challenges.

%\vspace{-.15in}
%\clearpage
%\small
\bibliographystyle{abbrv}
\bibliography{bib/immersive,bib/s&p,bib/metaverse,bib/ml}

\end{document}